\documentclass[aps,prl,twocolumn]{revtex4-1}
\usepackage{amsmath,hyperref}
\hypersetup{colorlinks=true,allcolors=blue}
\makeatletter
\def\@fnsymbol#1{$\, \flat$}
\makeatother

\renewcommand{\[}{\begin{equation}\begin{aligned}}
\renewcommand{\]}{\end{aligned}\end{equation}}

\def\tp{\theta^p}
\def\tq{\theta^q}
\def\ts{\theta^{\sigma}}
\def\tta{\theta^{\tau}}
\def\wdg{\wedge}

\def\mM{\mathcal{M}}

\def\mubar{\bar{\mu}}

\def\Rt{\tilde{R}}

\def\ep{\epsilon}

\begin{document}

\title{Duality in Einstein's Gravity}
\author{Uri Kol}
\email[E-mail address: ]{urikol@fas.harvard.edu}
\affiliation{Center of Mathematical Sciences and Applications, Harvard University, MA 02138, USA}

\begin{abstract}
We show that the Einstein equations in the vacuum are invariant under an $SO(2)$ duality symmetry which rotates the curvature 2-form into its tangent space Hodge dual. Akin to electric-magnetic duality in gauge theory, the duality operation maps classical solutions into each other. As an example, we demonstrate that the Kerr solution is non-linearly mapped by duality into Kerr-Taub-NUT.
\end{abstract}

\maketitle

\section{Introduction}

Electro-magnetic duality has been a key feature behind our understanding of Quantum Field Theory for over a century.
As was noted a long time ago by Heaviside \cite{heaviside1892xi}, the equations of the free electromagnetic field possess a remarkable invariance that rotates the electric and magnetic fields into each other.
To see this, consider the Maxwell action
\[\label{MaxwellAction}
S_{\text{Maxwell}} =\frac{1}{2e^2} \int _{X} \, F \wedge   \ast F,
\]
where $\ast F$ is the spacetime Hodge dual of the field strength $F$, which is defined by
\[\label{SpacetimeDualityOperation}
(\ast F)_{\mu\nu} \equiv \frac{1}{2} \ep _{\mu\nu\rho\sigma}F^{\rho\sigma}.
\]
The resulting equations of motion and the Bianchi identity
\[\label{MaxwellEqns}
\text{Maxwell's equations:} & \qquad &  d \ast  F &= 0,  \\
\text{Bianchi Identity:} & \qquad  & dF &=0,
\]
are invariant under the following $SO(2)$ duality rotation
\[\label{EMduality}
\begin{pmatrix}
F \\ \ast F
\end{pmatrix}
\longrightarrow
\begin{pmatrix}
F' \\ \ast F'
\end{pmatrix}
=
\begin{pmatrix}
+\cos \theta & +\sin \theta\\
-\sin \theta & +\cos \theta
\end{pmatrix}
\begin{pmatrix}
F \\ \ast F
\end{pmatrix}.
\]
For example, the field of a Coulomb electric charge transforms under \eqref{EMduality} into the field of a dyon (a particle with both electric and magnetic charges)
\[\label{mapDyon}
F^{\text{Coulomb}} \longrightarrow F^{\text{Dyon}}.
\]
Nevertheless, it was only in 1931 that Dirac \cite{Dirac:1931kp} provided a consistent realization of the field of a magnetic monopole.
The theory of magnetic monopoles was later extended by 't Hooft and Polyakov to the non-Abelian Georgi-Glashow model \cite{tHooft:1974kcl,Polyakov:1974ek}.

The concept of duality, however, is not merely an ordinary symmetry, but an equivalence between different descriptions of the same system.
Montonen and Olive \cite{Montonen:1977sn} suggested that the original Georgi-Glashow model is dual to a theory in which the fundamental degrees of freedom are replaced by the monopoles (which are solitons of the original theory).
Seiberg argued that two different supersymmetric non-Abelian gauge theories lead to the same non-trivial long distance physics \cite{Seiberg:1994pq}.
These dualities map the weak coupling region of the theory to the strong coupling region of its dual, and constrain the infrared physics, as was demonstrated by Seiberg and Witten \cite{Seiberg:1994rs,Seiberg:1994aj}.

In the following we will demonstrate that Einstein's theory of general relativity possesses an $SO(2)$ duality symmetry which is reminiscent of the electromagnetic duality \eqref{EMduality}.

\section{Gravitational Duality}

In the first order formalism, the Einstein-Hilbert action
\[\label{EHaction}
S_{EH}  =  \frac{1}{16 \pi G }\int_{\mM} \, \Rt _{ab}\wedge \theta^{a} \wedge \theta^{b}. 
\]
is expressed in terms of the curvature 2-form $R^{ab}$ and a non-coordinate basis of 1-forms $\theta^a$ in the flat tangent space. The curvature 2-form is the field strength of the spin connection $\omega^{ab}_{\mu}$
\[
R^{ab} = d \omega^{ab} + \omega^a{}_c\wedge \omega^{cb}
\]
and its components are given by the projection of the Riemann tensor onto the tangent space
\[\label{Riemann}
R^{ab} = \frac{1}{2}R^{ab}{}_{cd} \, \theta^c \wedge \theta^d .
\]
The non-coordinate basis of 1-forms in the tangent space is related to the coordinate basis in spacetime through the vielbeins $e^a{}_{\mu}$
\[\label{nonCoordBasis}
\theta^a = e^a{}_{\mu}dx^{\mu}.
\]
Note that we are using a notation in which tangent space indices are denoted by Latin letters $a,b,c$, while spacetime indices are denoted by Greek letters $\mu,\nu,\sigma$.

In writing the action \eqref{EHaction}, we have defined the \emph{tangent space} Hodge duality operation
\[\label{HodgeTangent}
\Rt _{ab}\equiv (\star R )_{ab} \equiv \frac{1}{2}\ep_{abcd}R^{cd},
\]
which is distinct from the \emph{spacetime} Hodge duality operation \eqref{SpacetimeDualityOperation}, and which is the main objective of this letter.

The Einstein equation and the Bianchi identity can be written using differential forms as
\[\label{GravityEqns}
\text{Einstein's equations:} \qquad & \Rt _{ab} \wedge \theta ^b &=0,\\
\text{Bianchi Identity:} \qquad & R _{ab} \wedge \theta ^b&=0.
\]
In components, and upon projection onto spacetime indices, these equations reproduce the familiar form of the Einstein equation $R_{\mu\nu}-\frac{1}{2}g_{\mu\nu}R=0$ and the algebraic Bianchi identity $R^\mu{}_{[\nu\rho\sigma]}=0$, respectively (see appendix \hyperref[appendixA]{A} for more details).

It is now evident that the set of equations in \eqref{GravityEqns} is invariant under the following $SO(2)$ duality rotation
\[\label{dualityOperation}
\begin{pmatrix}
R_{ab} \\ \Rt_{ab}
\end{pmatrix}
\longrightarrow
\begin{pmatrix}
R_{ab}' \\ \Rt_{ab}'
\end{pmatrix}
=
\begin{pmatrix}
+\cos \theta & +\sin \theta\\
-\sin \theta & +\cos \theta
\end{pmatrix}
\begin{pmatrix}
R_{ab} \\ \Rt_{ab}
\end{pmatrix}.
\]
We can further define self-dual and anti-self-dual components of the curvature 2-form
\[
R_{ab}^{\pm} = \frac{1}{2}\left(R_{ab}\mp i \Rt_{ab}\right)
\]
(note that $(R_{ab}^{\pm} )^* = R_{ab}^{\mp}$) which transform under the duality operation \eqref{dualityOperation} as
\[
R_{ab}^{\pm} \longrightarrow e^{\pm i\theta} R_{ab}^{\pm}.
\]
The $SO(2)$ duality rotation \eqref{dualityOperation} is a symmetry of the equations of motion and, consequently, it maps different solutions into each other, as will be demonstrated in the following section.

\section{Example}

As an example, let us consider the Kerr-Taub-NUT solution, whose metric is given in Plebanski coordinates \cite{Griffiths:2007ch} by 
\[\label{KTNmetric}
ds^2 = 
&+\frac{X}{p^2+q^2}\left(d\tau+q^2 d\sigma\right)^2
+\frac{p^2+q^2}{X}dp^2
\\
&-\frac{Y}{p^2+q^2}\left(d\tau -p^2 d\sigma\right)^2
+\frac{p^2+q^2}{Y}dq^2,
\]
with
\[
X &= a^2 -(p-\ell)^2,
\\
Y &= a^2 -l^2 -2mq +q^2
\]
(see appendix \hyperref[appendixB]{B} for more details about the Plebanski coordinates).
Here $m$ is the Schwarzchild mass, $\ell$ is the NUT parameter and $a$ is the spin parameter.

In order to evaluate the curvature 2-form and test the duality \eqref{dualityOperation}, we first define a coordinates basis in spacetime
\[
dx^{\mu} = (d\tau,d\sigma,dp,dq)
\]
and a non-coordinate basis
\[
\theta^a = (\theta^{\tau},\theta^{\sigma},\theta^{p},\theta^{q})
\]
in the flat tangent space, which we take to have a Lorentzian signature $\eta_{ab}=\text{diag}(-1,+1,+1,+1)$.
These two bases of 1-forms are related to each other by a basis of vielbeins \eqref{nonCoordBasis}, which we take to be
\[
e_a{}^{\mu} = \left(u^{\mu},v^{\mu},w_+^{\mu},w_-^{\mu}\right)
\]
with
\[\label{vielbeins}
u^{\mu} & = \frac{1}{\sqrt{(p^2+q^2)Y}} 
\left(+q^2,-1,0,0 \right),
\\
v^{\mu} & = \frac{1}{\sqrt{(p^2+q^2)Y}} 
\left(0,0,0,Y \right),
\\
w_{\pm}^{\mu} & = \frac{1}{\sqrt{2(p^2+q^2)X}}
\left(-p^2,-1,\pm X,0\right),
\]
(recall that $e_a{}^{\mu}$ is the inverse of $e^a{}_{\mu}$).
The metric \eqref{KTNmetric} is then reproduced by the standard formula
\[
g_{\mu\nu}=\eta_{ab}e^a{}_{\mu}e^b{}_{\nu}.
\]

Using this basis of vielbeins we can now evaluate the curvature 2-form (for example, by computing the Riemann tensor, projecting into the tangent space and using \eqref{Riemann}). We find that the self-dual and anti-self-dual parts of the curvature 2-form are given by
\[\label{KTNresult}
R_{ab}^+ = 
(R_{ab}^-)^* =
\frac{m+i\ell}{(q+i p )^3}\times H_{ab},
\]
where $H_{ab}$ is an anti-symmetric matrix of 2-forms, whose components solely depend on the fixed basis of 1-forms in the flat tangent space
\[
H_{12}&=-i H_{34} = \ts\wdg\tta+ i\, \tq \wdg \tp,\\
H_{13}&=+i H_{24} = \frac{\tta\wdg\tp+i\,\tq\wdg\ts}{2},\\
H_{14}&=-iH_{23}= \frac{\tta\wdg\tq-i\,\tp\wdg\ts}{2}.
\]
The simplicity of the expression in \eqref{KTNresult} for the curvature 2-form is striking.
While it represents the exact non-linear result that contains all the information about the curvature, the dependence on the parameters $m$ and $\ell$ turns out to be linear!
This is a special feature of the Plebanski coordinates and of the basis of vielbeins that we chose to work with \eqref{vielbeins}.

We can further define complex mass parameters
\[
\mu \equiv m+i \ell,
\qquad
\mubar \equiv m-i \ell,
\]
in terms of which the self-dual and anti-self-dual parts of the curvature 2-form are respectively holomorphic and anti-holomorphic functions that transform under duality \eqref{dualityOperation} as
\[
R_{ab}^{+}(\mu) &\longrightarrow  R_{ab}^{+}{}'(\mu) =e^{+ i\theta}R_{ab}^{+}(\mu)=R_{ab}^{+}(\mu '),
\\
R_{ab}^{-}(\mubar) &\longrightarrow  R_{ab}^{-}{}'(\mubar) =e^{- i\theta}R_{ab}^{-}(\mubar)=R_{ab}^{-}(\mubar '),
\]
with
\[\label{MassTransformation}
\mu ' \equiv e^{+i\theta} \mu,
\qquad
\mubar ' \equiv e^{-i\theta} \mubar
 .
\]
In other words, the transformation of the curvature 2-form amounts to a transformation of the mass parameters.

In terms of the real mass parameters the transformation \eqref{MassTransformation} reads
\[\label{massTransform}
\begin{pmatrix}
m' \\ \ell '
\end{pmatrix}
=
\begin{pmatrix}
+\cos \theta & -\sin \theta\\
+\sin \theta & +\cos \theta
\end{pmatrix}
\begin{pmatrix}
m \\ \ell
\end{pmatrix}.
\]
The complete curvature 2-form is the sum of both its self-dual and anti-self-dual parts
\[
R_{ab} =  R_{ab}^+ +  R_{ab}^-
\]
and it transforms under the duality operation \eqref{dualityOperation} as
\[
R_{ab}(m,\ell)\longrightarrow R_{ab}'(m,\ell) = R_{ab}(m ' ,\ell ').
\]

We see that under duality, the mass and the NUT parameter rotate into each other.
In particular, the curvature 2-form of the Kerr metric, which is obtained by setting the NUT parameter to zero, transforms under a general duality transformation \eqref{dualityOperation} into the curvature 2-form of the Kerr-Taub-NUT solution
\[\label{mapKTN}
R_{ab}^{\text{Kerr}} \longrightarrow R_{ab}^{\text{Kerr-Taub-NUT}}.
\]
The spinning version of \eqref{mapDyon}, namely a spinning Coulomb charge which is mapped by electromagnetic duality into a spinning dyon, is the field theory analogue of \eqref{mapKTN} (in certain instances the analogy becomes a concrete double copy map, see \cite{Monteiro:2014cda,Luna:2015paa,Arkani-Hamed:2019ymq,Huang:2019cja,Emond:2020lwi}).
Let us emphasize that our results are \emph{non-linearly} exact.

\section{Discussion}

Gravity in four spacetime dimensions can be interpreted as a gauge theory for the Poincar\'e group, in the sense that the vielbeins $e^a{}_{\mu}$ and the spin connection $\omega^{ab}_{\mu}$ can be viewed as the gauge fields for that group (see \cite{Gasperini:2017ggf} for example).
However, as argued by Witten \cite{Witten:1988hc}, the Einstein-Hilbert action does not take the form of a renormalizeable action in gauge field theory and therefore we cannot hope that four dimensional gravity would be a gauge theory in that sense.
Nevertheless, there are instances in which similarities arise and field theory techniques are still applicable in gravity. As an example, we have seen in this letter that the Einstein equations and the Bianchi identity are invariant under an $SO(2)$ duality symmetry, which is reminiscent of electromagnetic duality in field theory. A crucial difference, however, between duality in field theory and in gravity is that the former acts on the curved \emph{spacetime} indices while the later acts on the flat \emph{tangent space} indices.

The natural action of duality in the flat tangent space stems from the interpretation of Einstein's gravity as a theory of a curvature 2-form $R^{ab}$ on a \emph{fixed} basis of 1-forms $\theta^a$ in the flat tangent space.
Once a curvature 2-form is given, by means of dualisation or otherwise, the "potentials" $e^a{}_{\mu}$ and $\omega^{ab}_{\mu}$ can be deduced from it, and a spacetime manifold can be constructed by projection from the tangent space.
In a sense, gravity is emerging from a theory of a tensor valued 2-form for the Lorentz group in flat space.

Note that duality is a symmetry of the equations of motions and not of the Lagrangian.
Under duality, the Maxwell term \eqref{MaxwellAction} is mixed with the theta term, which is topological and therefore does not influence the equations of motion. Similarly, the Einstein-Hilbert action \eqref{EHaction} is mixed, under the duality operation \eqref{dualityOperation}, with the Holst term \cite{Holst:1995pc}, which is topological as well. In fact, the Holst term is even more trivial than an ordinary topological term - while it is not a total derivative, it simply vanishes identically by virtue of the Bianchi identity. A more complete treatment involves a larger phase space that includes torsion, in which the Holst term is part of the topological Nieh-Yan term, which is a total derivative. A detailed discussion will appear elsewhere.

Finally, let us comment on previous attempts to realize duality in general relativity. Most of these attempts are based on the following transformation of the Riemann tensor
\[\label{linearDuality}
R_{\mu\nu\rho\sigma} \longrightarrow \frac{1}{2}\ep_{\mu\nu\alpha\beta}R^{\alpha\beta}{}_{\rho\sigma}
\]
in linearized gravity (see \cite{Hull:2001iu,Henneaux:2004jw,Bunster:2006rt,Argurio:2009xr} for example).
However, the operation \eqref{linearDuality} is a symmetry of the linear theory \emph{only}. As shown in \cite{Deser:2005sz}, the operation above fails to be a symmetry already at first self-interacting, cubic, approximation of general relativity. This failure is also evident from the Kerr-Taub-NUT solution, which does not transform nicely under \eqref{linearDuality} beyond linear order.
On the contrary, the tangent space Hodge duality \eqref{HodgeTangent}, which is a symmetry, implies the following transformation
\[\label{NonLinearDuality}
R_{abcd} \longrightarrow \frac{1}{2}\ep_{abij}R^{ij}{}_{cd}.
\]
At linear order, the operation \eqref{linearDuality} coincides with the symmetry \eqref{NonLinearDuality}, but non-linearly these two operations are distinct. It would also be interesting to relate our results to the works of \cite{Alawadhi:2019urr,Banerjee:2019saj}.

\subsection{Acknowledgments}

I would like to thank Shing-Tung Yau for his valuable comments on the manuscript.


\bibliography{bibliography}
\bibliographystyle{apsrev4-2.bst}

\appendix

\section{Appendix A}\label{appendixA}

Here we show that the two equations in \eqref{GravityEqns} indeed reproduce the familiar form of the Einstein equation and the algebraic Bianchi identity. We refer the reader to \cite{Gasperini:2017ggf} for more details.

The explicit form of the first equation in \eqref{GravityEqns} is
$
\frac{1}{2}\ep_{abcd} R^{cd}\wdg \theta^b =
\frac{1}{4}\ep_{abcd} R^{cd}{}_{ij} \,  \theta^b \wdg \theta^i\wdg \theta^j =0.
$
Upon multiplying the components of the above 3-form by $\ep^{kbij}$ we arrive at
\[\label{eqnA}
\ep^{kbij}\ep_{abcd} R^{cd}{}_{ij}=0.
\]
Now we use the identity
\[
\ep^{dijk}\ep_{dabc} = - \delta ^{ijk}_{abc} \equiv 
-\det
\begin{pmatrix}
\delta^{i}_a & \delta^{j}_a & \delta^{k}_a \\
\delta^{i}_b & \delta^{j}_b & \delta^{k}_b \\
\delta^{i}_c & \delta^{j}_c & \delta^{k}_c \\
\end{pmatrix}
\]
to rewrite \eqref{eqnA} as
$
R^k{}_a - \frac{1}{2} \delta^k_a R = 0,
$
which, upon projection onto spacetime indices, reproduces the familiar form of the Einstein equation $R_{\mu\nu}-\frac{1}{2}g_{\mu\nu}R=0$.

The explicit form of the second equation in \eqref{GravityEqns} is
\[
\frac{1}{2}R^a{}_{bcd} \, \theta^{c}\wdg \theta^{d} \wdg \theta^{b} 
=
\frac{1}{12}R^a{}_{[bcd]} \, \theta^{c}\wdg \theta^{d} \wdg \theta^{b} 
= 0.
\]
The vanishing of the above 3-form components indeed reproduce, upon projection onto spacetime indices, the algebraic Bianchi identity $R^{\mu}{}_{[\nu\rho\sigma]}=0$.

\section{Appendix B}\label{appendixB}

The Kerr-Taub-NUT metric in the Boyer-Lindquist system of coordinates is given by \cite{Griffiths:2007ch}
\[\label{KTNBL}
ds^2 = &- f \left(dt +\Omega d\phi\right)^2
+\frac{\rho^2}{\Delta}dr^2
\\
&+\rho^2\left(d\theta^2 +\Sigma^2 \sin\theta ^2 d\phi ^2 \right),
\]
where
\[
\rho &= \sqrt{r^2 +(\ell+a \cos\theta)^2},
\\
\Delta &= r^2 -2mr+a^2-\ell^2,
\\
f&=1-\frac{2mr+2\ell(\ell+a \cos \theta)}{\rho^2},
\qquad
\Sigma = \sqrt{\frac{\Delta}{f\rho^2}},
\\
\Omega &= 2\ell(-\zeta+\cos\theta)-(1-\frac{1}{f})a\sin\theta^2.
\]
Here $\zeta = \pm 1$ correspond to semi-infinite Misner strings at $\theta=0$ and $\theta=\pi$, respectively. $\zeta = 0$ corresponds to an infinite Misner string on the entire axis $\theta=0,\pi$.

The Kerr-Taub-NUT metric in Plebanski coordinates \eqref{KTNmetric} is related to the Boyer-Lindquist form \eqref{KTNBL} by the following change of coordinates \cite{Griffiths:2007ch}:
\[
\tau &\longrightarrow t+ \frac{2a\ell(1-\zeta)-(a+\ell)^2}{a}\phi,
\quad
&\sigma &\longrightarrow -\frac{1}{a}\phi,
\\
p &\longrightarrow \ell +a \cos \theta,
\quad
& q &\longrightarrow r.
\]

\end{document}